# Universal Conductance Fluctuations of Topological Insulators

Shuai Zhang, Zhaoguo Li and Fengqi Song[†]

(National Laboratory of Solid State Microstructures, Collaborative Innovation Center of Advanced Microstructures, and Department of Physics, Nanjing University, Nanjing 210093, China)

**Abstract:** As an exotic quantum condensed matter, the topological insulator (TI) is a bulk- insulating material with a Dirac-type conducting surface state. Such dissipationless transport of topological surface states (TSSs) is protected by the time-reversal symmetry, which leads to the potential applications in spintronics and quantum computations. Understanding the topological symplectic transport of the Dirac fermions is a key issue to study and design the TI-based devices. In this review, we introduce the progress on the universal conductance fluctuation (UCF) of TSSs.

Firstly, we report the two-dimensional UCF phenomenon in TIs, and its topological nature is demonstrated based on the investigations of UCF by angle-varying, in-plane field tuning and scaling analysis. Secondly, we discuss the statistical symmetry of UCF in TIs. For a single TSS, the applied magnetic field will drive the system from a Gaussian symplectic ensemble into a Gaussian unitary ensemble. It results a $\sqrt{2}$ fold increase of the UCF amplitude. However, the experiment reveals a decreasing of the UCF amplitude of $\sqrt{2}$ times. This is contradictory to the theoretical prediction. Actually, there are two TSSs and they are coherently coupled to each other in TIs since the sample's thickness is shorter than its bulk dephasing length. This leads to a Gaussian orthogonal ensemble of the interface coupling system without an external field. In such situation, the UCF amplitude will decrease by $\sqrt{2}$ times with the field increasing. It is consistent with the experimental results. Finally, the other progress on UCFs is also discussed.

**Keywords:** topological insulators, quantum transport, universal conductance fluctuations, Berry phase

† Corresponding authors. E-mail: songfengqi@nju.edu.cn

# 1. Introduction

As an exotic quantum condensed material, the three-dimensional (3D) topological insulator[1,2] (TI) is bulk-insulating but there are conducting carriers within its surface. The surface carriers are known as Dirac fermions due to satisfying the Dirac equation. In momentum space, the dispersion relation of the Dirac fermions forms a linear Dirac cone, which runs through the bulk gap. Topological surface states (TSSs) contribute to the nontrivial topology of the band structure of the crystal and are protected by the time-reversal symmetry so that TSSs will not be destroyed by non-magnetic atoms doping or oxidation. Meanwhile, Dirac fermions will get a π Berry phase, when they rotate around along the Fermi surface. It guarantees the dissipationless transport properties of TSSs. So TI has potential applications in spintronics and quantum computations.

In terms of materials[3] studying, dozens of materials are considered to be TIs by theoretical predictions, and were immediately confirmed by angle-resolved photoelectron spectroscopy (ARPES) and scanning tunneling spectroscopy (STS). Among these materials, the family of $Bi_2Se_3$[4,5] attracts a lot of experimental scientists' interest because of its simple band structure. However, it is difficult to suppress the excess bulk carriers. For which reason, $Bi_2Te_{3-x}Se_x$[6] and $Bi_{2-x}Sb_xTe_{3-y}Se_y$[7] are becoming more popular. In the study of the physical properties[8] of TIs, the exotic properties of TSSs can be revealed by the transport measurements, which bring benefit to the study on the electronic devices associated with TI, especially reveal the π Berry phase is a key issue in the topological transport phenomena. So far, transport experiments, such as Shubnikov-de Haas (SdH) oscillations[9,10], quantum Hall effect (QHE)[11], Aharonov-Bohm (AB) oscillations[12,13] and weak antilocalization (WAL)[14,15], show sensitive response to TSSs Berry phase. As an important mesoscopic quantum interference[16] effect, universal conductance fluctuation (UCF) shows its unique and intriguing perspective in terms of revealing quantum transport properties of TSSs.

In this review, we introduce the progress on the UCF of TSSs by reviewing the outcomes and existing problems with the hope to inspire future studying. This article focuses on the studies of UCF which were systematically explored by our group[17] since 2011, and generally summarizes other groups' work. We begin in Section 2 with a general introduction to UCF. Section 3 describes the conductance fluctuation (CF) phenomena found in TI bulk crystal.

Section 4 is devoted to the identification of the UCF features in TI nanoribbons and its dimensional characteristic. We also preliminary investigate its nature of topology. In Section 5, we discuss the statistical symmetry of UCF in TIs. Section 6 briefly reviews the latest progress of UCF. We conclude with a discussion of the progresses and open questions in Section 6.

**2. Overview of universal conductance fluctuation**

UCF was first observed in the magnetoresistance (MR) of small-size conductor. When in 1984, Webb's group[18] in IBM used a golden ring with the diameter of 280 nm and ring arm width of 45 nm to study the AB effect of ordinary metal ring. But its MR showed aperiodic CF instead of the expected periodic AB oscillations. This phenomenon was later referred to be UCF by P. A. Lee[19]. It is also a mesoscopic quantum interference effect. Then Webb *et al.* increased the diameter to 825 nm while maintaining ring arm width, and finally observed clear AB oscillations[20], which superimposed on the background of slow fluctuation. After the UCF theory was proposed[19], it was understood why AB oscillation was not observed in the Au ring with smaller diameter, which is due to the UCF amplitude exceeded AB oscillation when electrons transported diffusively in the ring arm. While in the larger Au ring, the lengthened diffusion path inhibited the UCF amplitude, and make AB oscillation clear. Following is a brief description of the physical mechanism of UCF.

For macroscale conductors, the relationship between the resistance and the size can be expressed as

$$R = \rho \frac{L}{S}, \qquad (1)$$

where $\rho$ is the resistivity of the conductor, $L$ and $S$ are the length and cross-sectional area. In order to compare the different conductivity among the conductors, size-independent resistivity $\rho$ or conductivity $\sigma = 1/\rho$ is used. If the macroscopic conductor is divided into the small pieces of the same shape, is equation (1) always true along with decreasing the size of the conductor? In fact, when the conductor size is the same order of magnitude with a characteristic length, equation (1) is no longer true. This characteristic length is the coherence length $L_\phi$, also known as dephasing length. In this case, the conductivity of the conductor is directly described by the conductance $G=1/R$. This phenomenon can be explained by the diffusive transport mechanism illustrated in Figure 1. As electrons in a conductor diffuse from the source to the drain, the diffusion path of electrons is extremely complex due to impurity scattering. There are many self-closing path forming, and even some electrons diffusive back to the source. In the diffusion region, electrons in a conductor can move along the different diffusion path, which is called Feynman path. Each Feynman path can be expressed by wave function $A_i = |A_i|e^{i\varphi_i}$, and the total conductance is determined by superposition of wave function of all the Feynman path. So the probability of electrons diffusing from the source to the drain can be wrote as the following formula,

$$P_{\text{SD}} = |\sum_i A_i|^2 = \sum_i |A_i|^2 + \sum_{i \neq j} |A_i||A_j|\cos(\varphi_i - \varphi_j). \tag{2}$$

The second term in the formula is a quantum interference term. Obviously, the conductance is proportional to the probability, $G \propto P_{SD}$. Consider an impurity ensemble, which has the same macroscopic parameters of material, shape, size, impurity type and impurity concentration, and they have similar average degree of disorder and different impurity configurations. If the sample size is large enough, the averaging of all the interference term caused by Feynman path will be zero. Then all the samples in the impurity ensemble would get the same conductance described by equation (1). As the sample size decreases, Feynman path of electron diffusion will be gradually reduced. When the size is in the same order with coherence length $L_\phi$, the interference term cannot be eliminated by the averaging of Feynman path. In this case, the quantum correction is important. Every sample in the impurity ensemble has different impurity configurations, which corresponding to different Feynman path, and the interference is also different. Therefore, the conductance $G$ of each sample varies from each other. With respect to its mean value $\langle G \rangle$, where $\langle \cdots \rangle$ represents the ensemble averaging, the CF $\delta G = G - \langle G \rangle$ shows some statistical distribution. But the root mean square (RMS) of $\delta G$ is an universal value[21],

$$\delta G_{\text{rms}} = [\langle (\delta G)^2 \rangle]^{1/2} \approx \frac{e^2}{h}. \tag{3}$$

It has nothing to do with material, size and degree of disorder, so it is called UCF. Theoretical studies show that $\delta G_{rms}$ only has little dependence on the shape and dimensions of the sample. If the size of the conductor is $L^d$, where $d$ is the dimension of the system, the theory proves that $\delta G_{rms} = c \cdot e^2/h$ when the temperature is zero[21]. $c$ is a constant, and $c = 0.729$ when the system is quasi-one-dimensional, $c = 0.862$ for 2D and $c = 1.088$ for 3D. Here the spin degeneracy of electron is taken into account.

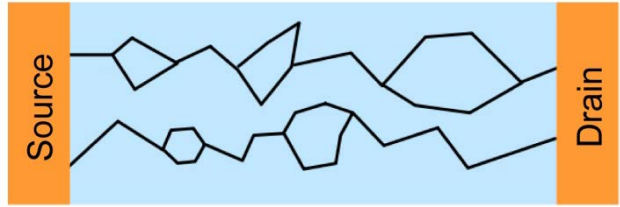

Figure 1. Electrons diffused transport in a metal.

As a kind of quantum interference, UCF is universal only when the system is coherent, i.e., $L < L_\phi$. If $L > L_\phi$, the size-averaging effect of different coherent regions will suppress $\delta G_{rms}$. The influence of temperature on UCF is more significant. Firstly, a finite temperature will affect the distribution of electrons at the Fermi level ($E_F$). Secondly, $L_\phi$ will decrease with temperature rising. It indirectly leads to size-averaging effect. Finally, thermal diffusion of electrons can also cause the loss of coherent. The characteristic length of the thermal diffusion is $L_T = \sqrt{\hbar D/k_B T}$, where $D$ is diffusion constant. In a word, the competition among $L$, $L_\phi$ and $L_T$ makes the transport of metallic conductors divided into several regions, and $\delta G_{rms}$ has different scaling law in every region[16].

With the presence of external magnetic field, each Feynman path will append a magnetophase, which is proportional to $\int \boldsymbol{A} \cdot d\boldsymbol{l}$. Because of the difference of phase of each Feynman path, the superposition of all Feynman paths leads the conductance oscillating randomly along with magnetic field. The special magnetoconductance (MC) fluctuations are corresponding to the impurity configurations in sample. As long as the impurity configurations do not change, the CF patterns are repeatable. Once it changes, for example, thermal diffusion will drive a part of impurities changing their position if the temperature of sample comes back to room temperature, the corresponding CF patterns will also change[22]. So, the patterns of UCF are known as magneto-fingerprint of the sample. In experiments, measuring MC fluctuation, i.e., $\delta G$-$B$, is a main way to study UCF. If follow the above definition of UCF, we need to measure such a large number of samples that the statistical $\delta G_{rms}$ can be obtained. And it should be ensured that all the samples have the same size, shape, degree of disorder and other parameters. However, this cannot be achieved in our experiments. When the phase accumulated along a Feynman path exceeds $h/e$ with the increase of the magnetic field, the conductance before and after the phase exceeding $h/e$ are no longer associated with each other. They correspond to different impurity configurations. Therefore, we can simulate all the impurity configurations, which traverse the entire ensemble. Quantitatively, it is the calculation of the autocorrelation function[21] of $\delta G$-$B$,

$$F(\Delta B) = \langle \delta G(B) \delta G(B + \Delta B) \rangle. \quad (4)$$

Obviously, $\delta G_{\mathrm{rms}} = [F(0)]^{1/2}$. Define the half high width of $F(\Delta B)$ as correlation field $\Delta B_c$, i.e., $F(\Delta B_c) = F(0)/2$. Phase-coherence length $L_\phi$ can be estimated[23] by $\Delta B_c L_\phi W \simeq \gamma h/e$, where $W$ is the width of the sample, and $\gamma$ is a constant. If $L_\phi \gg L_T$, $\gamma \simeq 0.95$. Or $\gamma \simeq 0.42$ when $L_\phi \ll L_T$.

In addition to MC fluctuation, UCF could be exhibited by other ways. For example, as the $E_F$ varying with gate voltage, it also shows CF, i.e., $\delta G$-$V_G$[24]; The voltage $\Delta V$ applied between the source and the drain would make the impurity potential attaching $-e\Delta V x$ potential, i.e., $\delta G$-$I$[25]; If there are two or more metastable states for a part of impurities in the conductor, then the quantum tunneling among the metastable states can still happen at low temperature, and the conductance fluctuates with time, i.e., $\delta G$-$t$[26]. All of the phenomena have fluctuation amplitude of $e^2/h$. It is the ergodic symmetry[19, 27] of UCF. UCF amplitude is also related to statistical symmetry[28]. It will be discussed in Section 5.

### 3. Conductance fluctuations in macroscopic Bi$_2$Se$_3$ crystals

Ong's group[29] observed the anomalous CFs in large non-metallic topological insulator Bi$_2$Se$_3$ bulk crystal firstly. As shown in Figure 2(a-c), as the smooth MR background was subtracted, the characteristics of CF at different temperature are almost the same. And at the lowest temperature, while changing the scanning direction of the magnetic field, the MC fluctuations are almost exactly the same. With the increase of temperature, the $\delta G_{rms}$ decreases gradually. All these features are consistent with the theoretical prediction of UCF. However, it is generally believed that UCF occurs only when the size of sample is the same order of magnitude with coherence length. It was the first time that UCF was observed in such a large bulk sample (2 mm×2 mm×50 μm). The RMS of the CF amplitude was calculated, $\delta G_{rms}$ = 5.9 $e^2/h$. According to the scaling law of UCF theory, the UCF amplitude of this sample can be estimated to be in the range of 0.01 to 0.05 $e^2/h$. But the $\delta G_{rms}$ observed in this $Bi_2Se_3$ crystal is nearly 500 times larger than that expected from the UCF theory, and the reason still remains elusive.

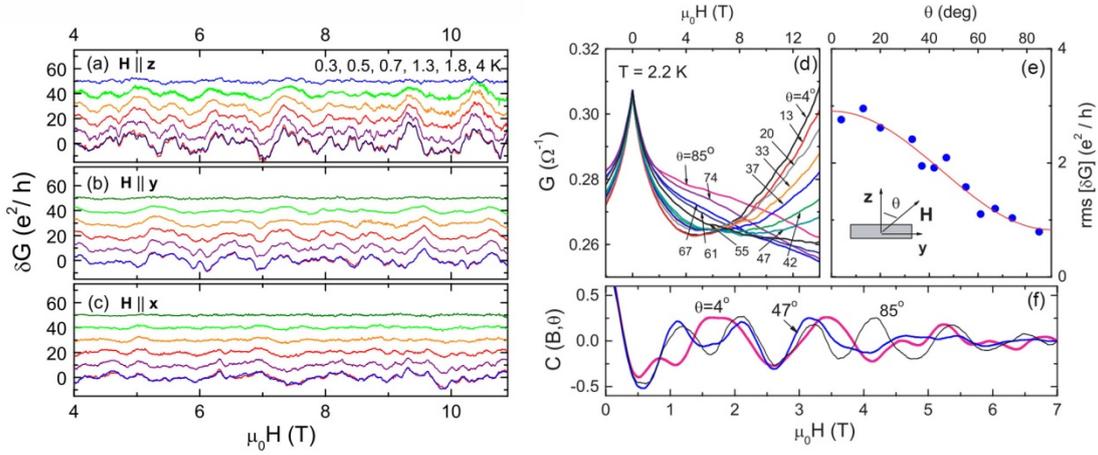

Figure 2. Conductance fluctuation phenomena of $Bi_2Se_3$ bulk. (a-c) show the $\delta G$-$B$ curves at different temperatures while the field aligned with three crystal axes. (d) The MC curves at different angle $\theta$. (e) $\delta G_{rms}$ as a function of $\theta$. (f) The self-correlation functions at different angles, where function $C(B)$ defined as $C(B)=F(B)/F(0)$. Adapted from Ref. 29.

They also investigated the CF behavior in different magnetic field, which showed in Figure 2(d-f). Figure 2(e) showed the relationship between CF $\delta G_{rms}$ and angle $\theta$. It can be described by the function of $(a + b\cos^2\theta)\, e^2/h$, where the first term is spin related, and the second term represents the orbit coupling effect. Angle dependent CFs phenomenon shows that the spin of carries plays an important role in magnetotransport. The TSS contribution in these phenomena needs to continue to study.

Later, Matsuo et al.[30] in Kyoto University studied the magnetotransport of $Bi_2Se_3$ grown by molecular beam epitaxy (MBE). They observed that $\delta G_{rms}$ is in the order of 0.01 $e^2/h$. It is in line with UCF theory. Because these samples were doped heavily by bulk carriers, it is not conducive to the study of TSS transport properties.

**4. Two-dimensional UCF in Bi$_2$Te$_2$Se nanoribbons**

Our group studied UCF in mesoscopic TI device systematically on the basis of obtaining bulk insulator TI crystal. Several experimental evidences on the nature of UCF of TSS were obtained. Then we would present an overview of the progress in this area.

We prepared bulk insulating Bi2Te2Se crystal. Microflakes were exfoliated and deposited on the 300-nm-SiO$_2$/Si substrates[31]. Then the Au electrodes were applied by photolithography and electron beam evaporation (EBE), and transport measurements were carried out using physical property measurement system (PPMS).

**4.1 Identifying the UCF features in Bi$_2$Te$_2$Se nanoribbon**

Figure 3 shows magnetotransport characteristics of a typical Bi$_2$Te$_2$Se nanoribbon. Figure 3(a) shows the temperature dependent resistance, i.e. *R-T* curve. At high temperature (*T* > 60 K), the resistance increases when temperature decreases, which shows the transport features of thermal activated carriers. The right inset shows the fitting of Arrhenius equation, $R(T) = R_0 e^{\Delta/2k_B T}$, which obtains an energy gap $\Delta$= 4.9meV. So the $E_F$ is located in the bulk band gap. At low temperature (*T* < 60 K), the resistance decreases as temperature decreases. It exhibits metallic transport characteristic, and this is supposed to be the surface state (SS) contribution. So when $k_B T \ll \Delta$, transport would be dominated by SS as the magnetic field is regulated.

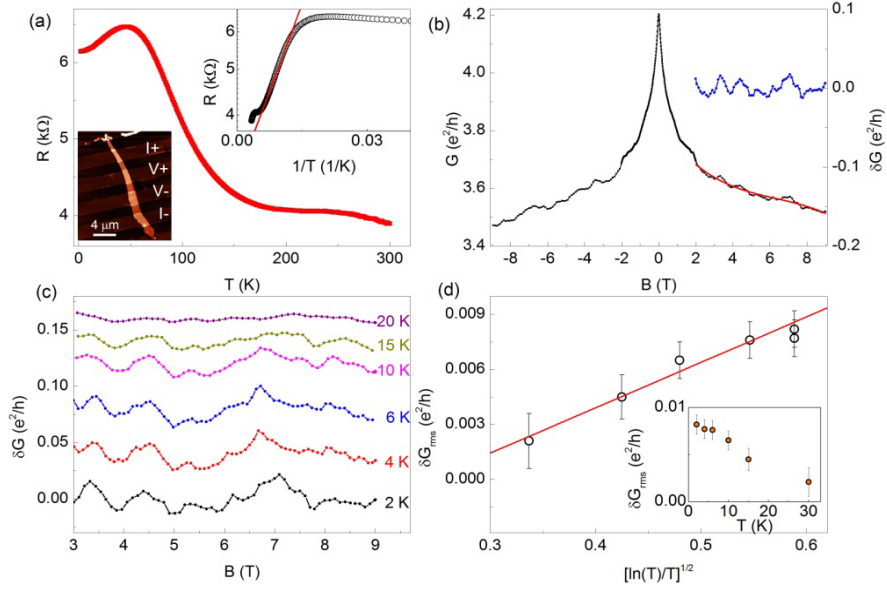

Figure 3. Analysis of UCF data. (a) The R-T curve, the left inset is the AFM image and the right inset is the Arrhenius fitting of R-T curve. (b) The MC curve at T = 2 K, the red curve is the polynomial fitting result. The blue curve is the conductance fluctuation curve after subtracting background. (c) The $\delta G$-$B$ curves at various temperatures. (d) Temperature dependence of $\delta G_{rms}$.

Figure 3(b) shows the MC curve at 2K as magnetic field perpendicular to the surface, i.e. $\theta = 0°$, where $\theta$ is defined as the angle between the magnetic field direction and the normal direction of the sample surface. It is obvious that there is a sharp peak around the zero field, known as the WAL effect. It is a response to the π Berry phase of TSS. In the high-field range, a series of random CFs can be observed. After subtracting the smooth background curve (red curve in Figure 3(b)], the aperiodic CF pattern can be observed clearly ($\delta G$-$B$, blue curve in Figure 3(b)). Figure 3(c) shows the $\delta G$-$B$ curves at different temperatures, and the CF patterns can be repeated at each temperature. Moreover, with the increase of temperature, the CF amplitude decreases gradually, which rules out the possibility that the CF comes from the thermal noise. First, the thermal noise will become more significant as the temperature rises. Another reason is that thermal noise is completely random and irrelevant at different temperatures. They cannot present the same fluctuation features. And at the same temperature, we can obtain the same $\delta G$-$B$ curve while changing the scanning direction of magnetic field.

This was observed in another sample and did not show in this paper. So such strong random CFs must be originated in quantum effect. UCF theory can describe the phenomenon.

Another feature of UCF is that the $\delta G_{rms}$ is about a quantum conductance $e^2/h$. But it requires the sample size is less than the coherence length ($L < L_\phi$), that is the whole sample is in one coherence zone. However, the $L_\phi$ of our sample is around 100 nm, which can be calculated from the autocorrelation function $F(\Delta B)$ of $\delta G$-$B$ curve, and the sample size is in the order of micron. This leads to a sample with the length $L$ and width $W$ contains $N = L \times W/L_\phi^2$ coherent zones. We always observe a smaller $\delta G_{rms}$ because of the averaging effect. For example, the $\delta G_{rms}$ of the sample shown in Figure 3 at 2K is 0.008 $e^2/h$. The temperature dependent $\delta G_{rms}$ is shown in Figure 3(d). We assume that the system is two-dimensional (2D), which would be proved true later. Then the scaling law of $\delta G_{rms}$ with temperature can be got, $\delta G_{\text{rms}} \propto (\ln T/T)^{1/2}$, according to 2D UCF theory[16]. It can be seen in Figure 3(d) that the experiment data are in good agreement with theory.

### 4.2 Identification of 2D UCF feature in Bi$_2$Te$_2$Se nanoribbon

The above scaling law of $\delta G_{rms}$ with temperature implies the UCF is 2D. And it can be further confirmed[32] by measuring MC with variable angle $\theta$, which is the main point of this section.

As stated earlier, UCF magneto-fingerprint of the system is caused by phase interference of different Feynman paths. For a strictly 2D system, when there exists an angle $\theta$ between the plane and the magnetic field (shown in Figure 4(a)), the phase accumulated by each Feynman path is only related to the plane normal component, that is $\int \boldsymbol{A} \cdot d\boldsymbol{l} \propto B\cos\theta$. Then, the quantum interference of all Feynman path also depends only on $B\cos\theta$. So are the CF peaks. Figure 4(b) exhibits the $\delta G$-$B$ curve with different $\theta$. Along with the increase of $\theta$, conductance peaks move monotonously and broaden in the direction of high field. We marked three typical conductance peaks (p1, p2, p3) position changing with $\theta$. And it was portrayed as a function. It obviously matches the $1/\cos\theta$ law (Figure 4(c)). Thus the 2D nature of UCF is revealed. It is the first time to observe the 2D UCF phenomenon in TI. The 2D UCF were then observed in $Bi_2Se_3$[33], $Bi_{1.5}Sb_{0.5}Te_{1.7}Se_{1.3}$[34] and Kawazulite[35] (a kind of minerals which composition is approximately $Bi_2(Te,Se)_2(Se,S)$) by other groups.

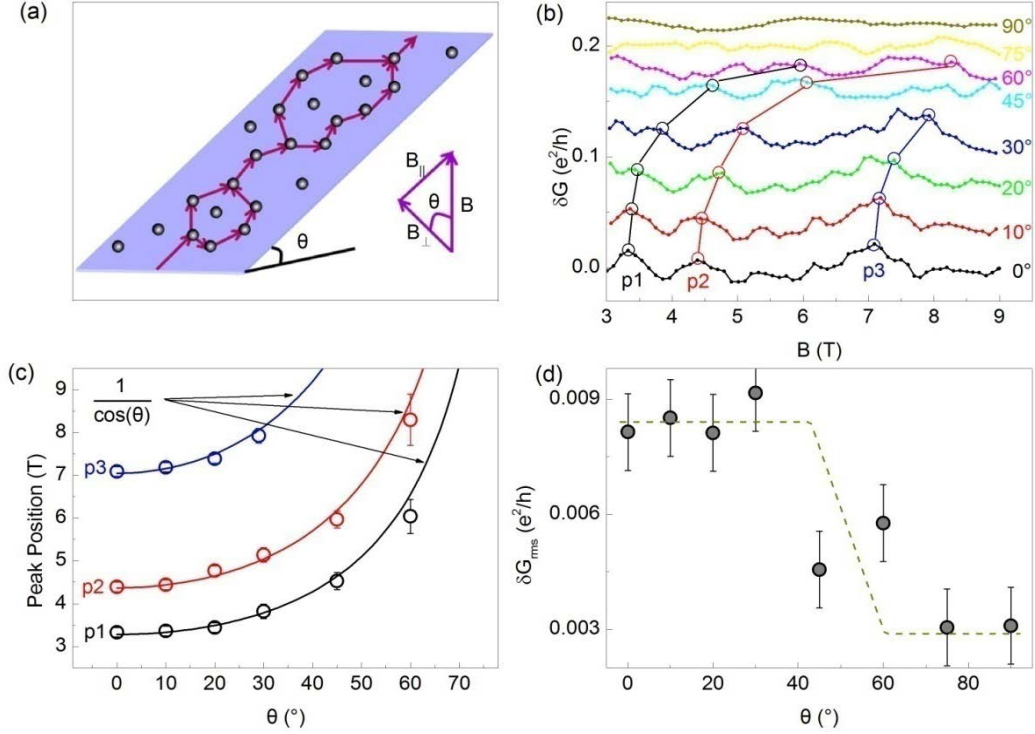

Figure 4. The two-dimensional (2D) UCF nature. (a) The schematic diagram of 2D UCF. (b) The *δG-B* curves at 2 K. The black, red and blue circle-marked lines respectively show the similar features, namely p1, p2 and p3, in all the *δG-B* curves. (c) The positions of the UCF features plotted against *θ*. The black, red and blue data are from those of p1, p2 and p3 in (b), respectively. The solid curves are the 1/cos*θ* fitting. (d) *θ* dependent *δG$_{rms}$*. The dashed curve is for eye guiding.

As we all know, the electrical transport of 3D TIs is mainly composed of 3D bulk state and 2D SS. In the $Bi_2Te_2Se$ system, although $E_F$ is in the gap, the impurity state caused by bulk impurities may form impurity band within the gap, which contributes to the conductance[6, 36, 37]. However, 3D bulk UCF would not show the $1/\cos\theta$ law. Therefore, the 2D UCF should come from 2D SSs. In addition, *δG$_{rms}$* with various *θ* also supports this conclusion (Figure 4(d)). Because the electrons of 3D system can be controlled by the magnetic field in every direction, its *δG$_{rms}$* is insensitive to *θ*. But for 2D system, electrons will not be affected by magnetic field when the magnetic field is in plane, so *δG$_{rms}$* is dependent seriously on $\theta$. As shown in Figure 4(d), $\delta G_{rms}$ is almost not affected by $\theta$ when $\theta <$ 45°, which is the influence of SS dominated transport. While $\theta > 45°$, SS contribution reduces; this leads $\delta G_{rms}$ decreases rapidly to about 0.003 $e^2/h$. When the field is parallel to the axial of nanoribbon, i.e. $\theta = 90°$, SS no longer contribute to MR, and the MR should be contributed by bulk. So the UCF at $\theta = 90°$ should come from bulk (Figure 4(b)). Such a weak UCF may also be caused by the noise of measurement system. WAL measured at various $\theta$ shows 2D transport properties, which further confirms the conclusion that 2D SS is dominated in the sample.

We proved the 2D nature of transport features (UCF and WAL) by means of regulating $\theta$. And the 2D features are due to 2D SS. The premise of this conclusion is that the bulk is in 3D state. It needs to consider the coherence length. Here are two methods. One is to estimate by UCF. We could believe the weak CF comes from the bulk when $\theta = 90°$. A coherence length of 12 nm can be deduced according to the classic 3D averaging effect. It is less than the sample thickness of 62nm. Thus the bulk state is 3D. The other way to calculate the coherence length is fitting WAL[38]. Analyzing the MR at $\theta = 90°$ by 3D WAL theory can obtain the coherence length of about 34nm. It is also less than the thickness. So we can say that the 2D UCF and WAL features are caused by SS.

We measured many other samples[39]. And the 2D UCF and WAL feature were observed in all of the samples. But the samples do not all have strict 3D bulk state. The bulk coherence length is about 60 nm in most of them, and so is the thickness. The thickness of some samples is even less. For example, Figure 5 shows the transport behavior of a 47 nm thick sample. Its bulk coherence length is 50 nm. This means that the bulk state is in the cross region of the 3D to 2D. The bulk carries of this situation can also exhibit quasi-2D properties, which will disturb the identification of SSs. So other ways to rule out the 2D bulk contribution are needed. The in-plane magnetic field regulation is an appropriate choice[39].

As shown in the insert of Figure 5(d), we applied both the vertical ($B_\perp$) and in-plane magnetic field ($B_\parallel$) for the sample at the same time. $B_\parallel$ will not influence the transport properties of SSs but suppress the coherence of bulk carries. So we can scan $B_\perp$ while keeping $B_\parallel$ fixed. By adopting UCF signals from $G$-$B_\perp$ curves, the bulk state influence can be known. According to UCF theory[16], $\delta G_{\rm rms} \propto L_\phi^{(4-d)/2}$ when $L_\phi \ll L$, where $d$ is the dimension of system. The $\delta G_{rms}$ is sensitive to coherence length. Therefore, if the quasi-2D bulk contributes to UCF, its UCF amplitude will reduce with the increase of $B_\parallel$. While UCF of SSs will not be affected by $B_\parallel$.

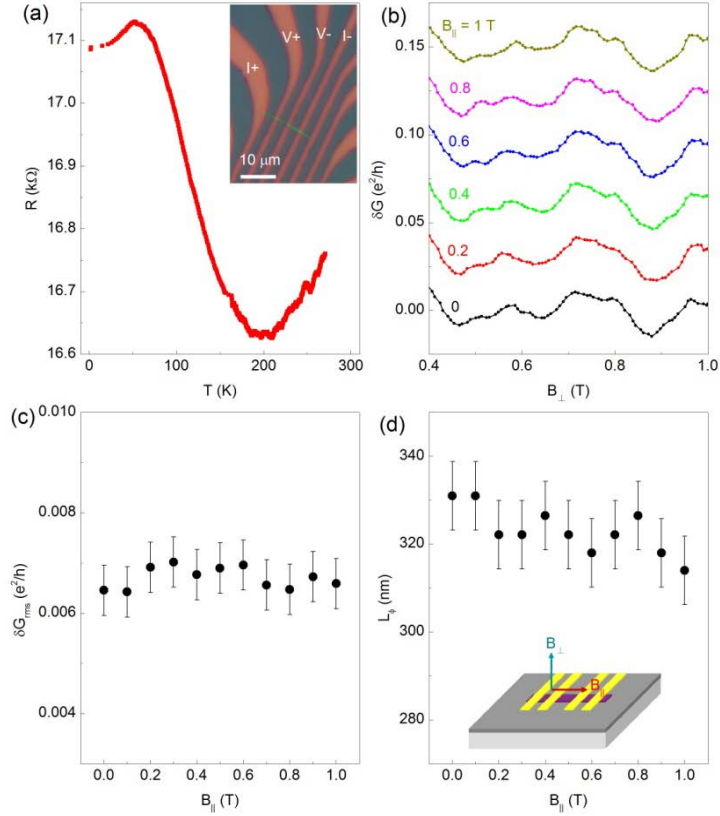

Figure 5. Tuning UCF by a parallel field ($B_\parallel$). (a) The R-T curve, the inset is the optical image of sample. (b) The $\delta G$-$B$ curves at various $B_\parallel$. $B_\parallel$ dependence of $\delta G_{rms}$ (c) and $L_\varphi$(d). The inset of (d) shows the magnetic field configuration.

Figure 5(b) shows the $\delta G$-$B_\perp$ curves of the 47 nm thick sample at different $B_\parallel$. Its $\delta G_{\rm rms}$ and $L_\phi$ have been depicted as a function of $B_\parallel$ in Figure 5(c) and 5(d). $\delta G_{\rm rms}$ is almost unaffected by $B_\parallel$, which is in line with the expectation. However the $L_\phi$ of bulk carriers drops from 50 nm to 17 nm while $B_\parallel$ increasing from 0 to 1 T, which means a significant change in $\delta G_{\rm rms}$. It is inconsistent with the experiment. Thus we confirm the SS dominated quantum transport in $Bi_2Te_2Se$ nanoribbon once again.

Now we proved the transport of $Bi_2Te_2Se$ nanoribbon is dominated by SS by regulating tilted magnetic field and in-plane field. Although bulk carries also contribute a certain conductance, the two means of regulation mentioned above both proved the correction to transport is extremely weak. That is way the conductance response caused by vertical magnetic field can be seen as the SS response directly. Ultimately, the TSS transport properties can be obtained.

**4.3 Indications of the topological nature of 2D UCF**

Through the above discussion, we can be sure that the UCF in $Bi_2Te_2Se$ nanribbon comes from the surface carriers, but not the bulk carriers. But we cannot say the UCF originates from TSSs yet, because there is two-dimensional electron gas (2DEG) in TI surface. Here we try to distinguish them.

The trivial 2DEG on the TI surface is caused by the bulk band bending near the surface[40]. On the interface between TI and vacuum, the Fermi levels on the both sides of the interface do not match each other, and it generates electrostatic potential. So the conduction band and valence band move downwards and there forming a quantum well structure near the interface. This quantum well can blind the dopant of the bulk and the environment, for example being exposed in the air. The doped carriers will form trivial 2DEG. And the strong spin orbit coupling (SOC) makes significant Rashba splitting on 2DEG, which separates the spin up and spin down subband. Surface electrostatic potential not only reconstructs the bulk band and 2DEG but also makes the Dirac cone of TSS move down further, which leads to the Dirac point (DP) away from $E_F$, making the detection of transport behavior near the Dirac point more difficult. The trivial 2DEG on TI surface has been observed by APRES experiment[40, 41]. With the increase of exposing time, $E_F$ moves towards bulk band. And the Rashba splitting of 2DEG will become more significant[42, 43]. This indicates that the environment doping is an important source of the formation of 2DEG. Transport experiment can also reveal the existence of 2DEG on TI surface[15, 44, 45]. All in all, the 2DEG are proved to exist on the TI surface by ARPES, transport experiment and theoretical calculation. It challenges us to identify the origin of 2D UCF.

Here, we propose that the origin can be determined by measuring the RMS of UCF. According to the UCF theory of TSS developed recently[46, 47], we are able to know UCF amplitude, $\delta G_{rms}$= (0.43-0.54) $e^2/h$. But the UCF amplitude of 2DEG is different, $\delta G_{rms}$= 0.86 $e^2/h$. So the origin can be clear by comparing the experimental data with the theoretical value. But as shown in Figure 2 to Figure 4, the UCF amplitude measured experimentally is in the order of 0.001 $e^2/h$. The reason is that our samples size is larger than the coherence length ($L > L_\phi$), and the averaging effect of different coherence zone makes the UCF amplitude reducing. Theoretical expectation is in the condition of $L < L_\phi$. Therefore, we need to take appropriate scaling law, and derive the UCF amplitude of single coherence zone, which we call intrinsic UCF amplitude, from the scaling law. Then we can compare them.

Considering the classic self-averaging effect, statistical averaging of $N = L \times W / L_\phi^2$ coherence zones makes intrinsic UCF amplitude inhibited by $1/\sqrt{N}$ times. Due to the theoretical calculation result is the conductivity of the system, we need to take the 2D conductivity $G_\Box = G \cdot (L/W)$. As a result, the relationship between the UCF amplitude $\delta G_{rms}$ measured by experiment and the intrinsic UCF amplitude $\delta G_{rms}^\Box$ is

$$\delta G_{\text{rms}} \simeq \beta \frac{\delta G_{\text{rms}}^\Box}{\sqrt{N}} \cdot \frac{W}{L} = \delta G_{\text{rms}}^\Box \cdot \beta \frac{L_\phi W^{1/2}}{L^{3/2}}, \qquad (5)$$

where $\beta$ is a constant and related to the symmetry of the system. Here take[48] $\beta = 1/2\sqrt{2}$.

In this work, we have measured 14 samples. The length $L$ and width $W$ of samples can be gained from atomic force microscopy (AFM) measurement. $\delta G_{rms}$ and $L_\phi$ can be extracted from $\delta G$-$B$ curves. And we can know the intrinsic UCF amplitude $\delta G_{rms}^\Box$ by equation (5). It is shown in Figure 6. We can find that the $\delta G_{rms}^\Box$ does not present a unified trend, but the distribution is spread. It can be explained by these reasons below. The impurity concentrations are different in each sample although these samples were exfoliated from the same crystal. Moreover, the Fermi levels of the samples are not located in the same position, which also affect the $\delta G_{rms}^\Box$. Lastly, equation (5) only considers the size-averaging effect but neglects the other effect, such as energy-averaging effect. So the simplified treatment of equation (5) could enlarge the experimental error, and cause the dispersion in Figure 6.

In theory, the UCF of a 2D geometry TSS can be described[49] as

$$\delta G_{\text{rms}} = \frac{e^2}{\pi^2 h} \left( \sum_{n_x=1, n_y=-\infty}^{+\infty} \frac{12 g_s g_v}{\left[n_x^2 + 4\left(\frac{L}{W}\right)^2 n_y^2\right]^2} \right)^{1/2}, \qquad (6)$$

where $g_s g_v = 1$ for TSS. This curve has been depicted in Figure 6 (solid line). With the increase of $L/W$, $\delta G_{rms}^{\square}$ decreases gradually. While $L/W \gtrsim 1$, $\delta G_{\text{rms}}^{\square}$ saturates to 0.37 $e^2/h$. And all of our samples meet the condition $L/W > 1$, so the size dependent $\delta G_{rms}^{\square}$ does not be observed. We still find that the experimental data are around the theoretical value of TSS but away from the prediction of trivial 2DEG, although there are many reasons causing the dispersion. This means that the UCF observed in $Bi_2Te_2Se$ nanoribbon is the response of TSS rather than 2DEG. Ultimately, we may determine the origin of UCF.

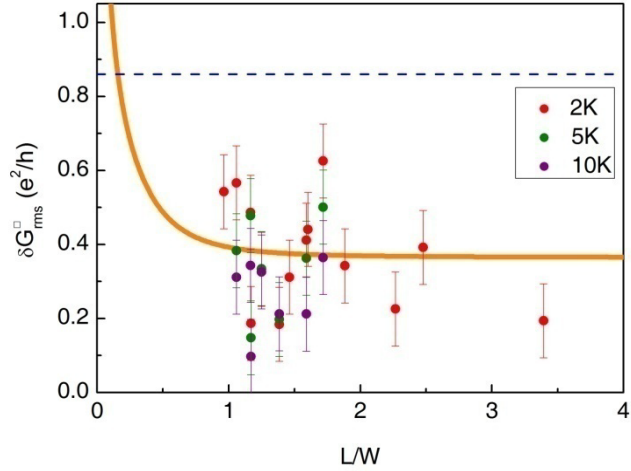

Figure 6. Intrinsic UCF amplitudes as a function of *L/W*. The closed-circle markers present the experimental data and the solid curve shows the theoretically expected values according to Eq. (6). The dashed line presents the expected values for a topologically trivial 2DEG.

As mentioned above, many factors, such as impurity concentration, the position of $E_F$ and averaging effect, will influence the $\delta G_{rms}^{\square}$ of SS. So we could preliminarily confirm the UCF originates from TSS by comparing the UCF amplitude. Further experimental study is needed to provide more reliable evidences. For example, in order to avoid the influence of averaging effect, the smaller samples ($L < L_\phi$) are needed; in order to avoid the different impurity concentration, we can fabricate the electrode with different distance on the same sample to study the size scaling law of UCF; in order to know the effect of $E_F$, gate voltage can be used. In addition, the lower temperature is necessary for reducing the thermal effect, and the longer coherence length means the larger UCF amplitude, which will reflect the contribution of TSS more clearly. In the next section, we will present a preliminary exploration in this regard.

## 5. Exploration of the statistics symmetry of TSS

The UCF of TI discussed above is based on a microscopic perspective[21], that is, consider the distribution of disorder impurities in the impurity ensemble and calculate the conductance of every impurity configuration by numerical simulation to get the $\delta G_{rms}$ and compare it with the experiment. In this section, we take a macroscopic perspective[50] to examine UCF. It is more concise than the microscopic perspective and can reflect the universality of UCF.

### 5.1 Statistics symmetry of UCF

In the macroscopic perspective, the conductance is determined by Landauer-Büttiker formula[51], i.e. transfer matrix $T$. Each impurity configuration corresponds to a transfer matrix $T$. In fact, the set of so many transfer matrixes can be seen as the set of random matrixes. Then the study of impurity ensemble becomes studying the set of transfer matrixes which satisfies certain conditions without considering the detail of impurity configuration. By limiting the random matrixes set to meet certain conditions which correspond to the characteristics of impurity ensemble, the statistic of random matrixes can exhibit CF features. This is the random matrixes theory[28, 52-59] of UCF. Here, the UCF feature is determined by the symmetry of random matrix set, and the detail of Hamiltonian will not affect the result. UCF amplitude given by random matrix theory is

$$\delta G_{\mathrm{rms}} \simeq c_d \frac{e^2}{h} \sqrt{ks^2/\beta}, \tag{7}$$

where $c_d$ is a coefficient related to the dimension $d$, $k$ is statistically independent eigenvalue sequence number of $tt^\dagger$ ($t$ is transmission matrix), $s$ is the degeneracy of eigenvalue sequence and $\beta$ is a constant related to the symmetry. Dyson introduced the Gaussian distribution of Hamiltonian and classified random with different symmetry into three ensembles which also called three universal symmetry classes[60]. Gaussian orthogonal ensemble (GOE) describes the system with both time reversal symmetry (TRS) and spin rotation symmetry. The Wigner-Dyson parameter $\beta = 1$ for this situation. Gaussian unitary ensemble (GUE) describes the TRS-broken system, and $\beta = 2$. Gaussian symplectic ensemble (GSE) describes the TRS system without spin rotation symmetry, i.e. there is the existence of SOC, and $\beta = 4$.

It can be seen that the systems with different universal symmetry classes have different $\delta G_{rms}$ from equation (7). The universal symmetry can be regulated by external parameters, so that the transition of $\delta G_{rms}$ in different universal symmetry can be observed. In general, the external magnetic field can break the TRS, which can cause the transition from GOE (or GSE) to GUE. The intensity of SOC can be controlled by gate voltage, leading GSE transforming into GOE. The change of $\delta G_{rms}$ at different universal symmetry can be calculated by equation (7), and the results are summarized in Table 1 (for simplicity, only the values of $\sqrt{ks^2/\beta}$ are shown in the table). Next, we discuss them respectively.

**Table 1.** UCF amplitude of different symmetry (Unit: $c_d e^2/h$)

| System | Case | $B = 0$ | $B_c < B < B_Z$ | $B > B_Z$ |
|---|---|---|---|---|
| Normal metal | Weak SOC | 2 | $\sqrt{2}$ | 1 |
|  | Strong SOC | 1 | $1/\sqrt{2}$ | $1/\sqrt{2}$ |
| Graphene | Near DP | 4 | $2\sqrt{2}$ | 2 |
|  | Far away DP | 2 | $\sqrt{2}$ | 1 |
| Single TSS | Near DP | 1 | $1/\sqrt{2}$ | $1/\sqrt{2}$ |
|  | Far away DP | 1/2 | $1/\sqrt{2}$ | $1/\sqrt{2}$ |

For disordered metal, we consider the system without SOC (or with very weak SOC) first. In the absence of magnetic field ($B = 0$), the system is GOE ($\beta = 1$). The eigenvalue degeneracy is only spin degenerate ($s = 2$, $k = 1$), the corresponding UCF amplitude $\sqrt{ks^2/\beta} = 2$. When the magnetic field exceeds the threshold $B_c = \phi_0/L_\phi^2$ ($\phi_0 = h/e$ is quantum flux), the TRS is broken. And the system becomes GUE ($\beta = 2$) with $s$ and $k$ unchanged, $\sqrt{ks^2/\beta} = \sqrt{2}$. Obviously, the UCF amplitude is reduced by $\sqrt{2}$ times in the transition from GOE to GUE. If the magnetic field is increased further, the system occurs significant Zeeman split and the spin is no longer degenerate ($s = 2 \rightarrow s = 1$). Then spin up and spin down eigenvalue sequences become independent ($k = 1 \rightarrow k = 2$) while the system remain in GUE, $\sqrt{ks^2/\beta} = 1$. As a result, for weak SOC metal system, with the increase of magnetic field, UCF amplitude will be reduced by $\sqrt{2}$ times when TRS is broken ($B > B_c$); and it is reduced again by $\sqrt{2}$ times when the magnetic field exceeds the Zeeman field ($B > B_Z$). It has been confirmed by experiment[61-65].

Then consider the disordered metal systems with strong SOC. The system belongs to GSE ($\beta = 4$) at zero field. Despite that SOC has caused spin degeneracy retired, the TRS is not broken and introduces Kramers degeneracy with double degeneracy ($s = 2$, $k = 1$), so $\sqrt{ks^2/\beta} = 1$. When the magnetic field exceeds $B_c$, which makes TRS broken, the system becomes GUE ($\beta = 2$) and Kramer degeneracy is retired ($s = 2 \rightarrow s = 1$) with $k$ unchanged, then $\sqrt{ks^2/\beta} = 1/\sqrt{2}$. When the magnetic field continues to increase, the UCF amplitude will not decrease any more. Therefore, the $\delta G_{rms}$ will be reduced by $\sqrt{2}$ times with the increase of magnetic field. It has also been proved true in experiment[66].

Furthermore, for ordinary metal, the system can change from GOE to GSE at zero field. At this time, the $\delta G_{rms}$ will be reduced by 2 times. There is still no experiment report on this change of UCF, though the intensity of SOI can be regulated by gate voltage effectively[67].

Next, we consider graphene. Here we still discuss by two kinds of situations, near DP and far away DP. When the $E_F$ is near DP, the scattering between two Dirac cone (inter-valley scattering) is suppressed, which makes pseudo-spin degenerated ($s_{valley} = 2$) and spin rotation symmetry broken. So graphene system of this situation is GSE ($\beta = 4$) at zero field. Taking into account the spin degeneracy ($s_{spin} = 2$) and Kramers degeneracy ($s_{Kramers} = 2$) of DP, the final degeneracy $s = s_{valley} \times s_{spin} \times s_{Kramers} = 8$ while $k = 1$. So there is $\sqrt{ks^2/\beta} = 4$. When the magnetic field exceeds and breaks TRS, the graphene system becomes GUE ($\beta = 2$). Then the Kramers degeneracy is lifted ($s_{\text{Kramers}} = 1$) while pseudo-spin and spin degeneracy still persists ($s_{\text{valley}} = 2$, $s_{\text{spin}} = 2$), and $s = 4$ and $k = 1$. So $\sqrt{ks^2/\beta} = 2\sqrt{2}$. If magnetic field is further increased so that the Zeeman spilt of electron spin occurs, then $s_{\text{valley}} = 2$, $s_{\text{spin}} = 1$, $s_{\text{Kramers}} = 1$, and $s = 2$ and $k = 2$. The system is still GUE, so $\sqrt{ks^2/\beta} = 2$.

When the $E_F$ of graphene is far away DP, the strong inter-valley scattering retire the pseudo-spin ($s_{\text{valley}} = 1$), and the system is spin rotation symmetric. Graphene system of this situation is GOE ($\beta = 1$) at zero field. Kramers degeneracy is retired ($s_{\text{Kramers}} = 1$) because $E_F$ is far away DP. So the degeneracy of the system is determined by spin degeneracy, $s = s_{\text{spin}} = 2$. For $k = 1$, $\sqrt{ks^2/\beta} = 2$. When $B > B_c$, TRS is broken, and the system becomes GUE ($\beta = 2$). For $s = 2$ and $k = 1$, we can know $\sqrt{ks^2/\beta} = \sqrt{2}$. When $B > B_Z$, spin degeneracy is lifted. There is $s = 1$, $k = 2$ and $\beta = 2$, so $\sqrt{ks^2/\beta} = 1$.

Let's sum up the statistical laws of UCF in graphene. First, Wherever the $E_F$ is, $\delta G_{rms}$ reduces by $\sqrt{2}$ times with the magnetic field increasing due to the TRS broken, and then reduces by $\sqrt{2}$ times further due to the lifting of the spin degeneracy. Second, no matter how strong the magnetic field is, i.e. TRS is broken or not and the Zeeman splitting is happened or not, the $\delta G_{rms}$ will reduce 2 times when the $E_F$ is regulated to be far away from the DP by gate voltage. It has been confirmed in experiment[48, 68, 69].

Last is the situation of TSS. In TI, there is only one Dirac cone for TSS, i.e. $s_{\text{valley}} = 1$, and SOC makes the spin always polarized, i.e. $s_{\text{spin}} = 1$. So the degeneracy of the system is determined by the Kramers degeneracy. It also the reason that the TSS system belongs to the GSE ($\beta = 4$) at zero magnetic field. When the $E_F$ is near DP, the Kramers degeneracy of the DP cause that $s = s_{\text{Kramers}} = 2$ at $B = 0$. For $\beta = 4$ and $k = 1$, $\sqrt{ks^2/\beta} = 1$ in this condition. Once $B > B_c$, the system changes to GUE ($\beta = 2$). The Kramers degeneracy is lifted due to the TRS broken, so $s = s_{\text{Kramers}} = 1$ and $k = 1$. Then $\sqrt{ks^2/\beta} = 1/\sqrt{2}$. If the magnetic field is further increasing, no other change would happen.

When the $E_F$ is far away from DP, there is no Kramers degeneracy, and so $s = 1$. At $B = 0$, the parameters are given by $\beta = 4$, $s = 1$ and $k = 1$, so $\sqrt{ks^2/\beta} = 1/2$. If $B > B_c$, there are $\beta = 2$, $s = 1$ and $k = 1$, so $\sqrt{ks^2/\beta} = 1/\sqrt{2}$. Still stronger magnetic field would not cause any other effect.

To summarize, there are following conclusions for a single TSS. First, when the $E_F$ is near DP, the $\delta G_{rms}$ will reduce $\sqrt{2}$ times with increasing magnetic field. Second, when the $E_F$ is far away from DP, the $\delta G_{rms}$ will be increased by $\sqrt{2}$ times with magnetic field increasing. This is different from other systems. Third, in the progress of the $E_F$ moving away from the DP, the $\delta G_{rms}$ will reduce 2 times at zero magnetic field, and keep unchanged when $B > B_c$.

### 5.2 Statistics symmetry of coexistence of two TSSs

In the previous section, we only discussed the statistics symmetry for a single TSS. In an actual system, there are two relative surfaces[15, 70, 71] in TI. Therefore, Table 1 is an ideal system, and it cannot be used in the experiments directly. Here, we will discuss the statistics symmetry of coexistence of two TSSs. There will be interactions between the two TSSs. So the coupling and decoupling situation should be considered, as well as the distance from DP. Therefore, there are totally for situations, as shown in Figure 7.

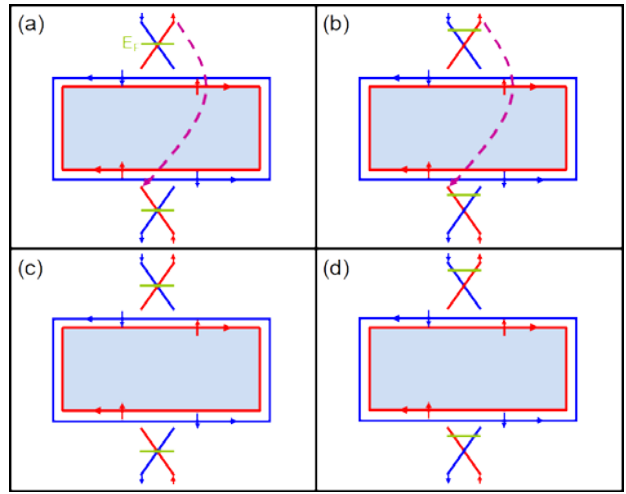

Figure 7. The schematic diagram of the two TSSs coupling/decoupling.

We consider the situation for the coupling TSSs and the $E_F$ near the DP firstly (Figure 7(a)). At $B = 0$, due to the coupling of two TSSs, it behaves as with only one cone, i.e. $s_{valley} = 1$. But it causes the spin degeneracy, i.e. $s_{spin} = 2$. So the spin rotation symmetry is established. Then the system is GOE ($\beta = 1$). With the Kramers degeneracy in DP, the total degeneracy $s = s_{valley} \times s_{spin} \times s_{Kramers} = 4$. Also $k = 1$, so $\sqrt{ks^2/\beta} = 4$. When $B > B_c$, the system belongs to GUE ($\beta = 2$). The Kramers degeneracy is lifted ($s_{valley} = 1$), but the coupling still exists ($s_{spin} = 2$). With $s = 2$ and $k = 1$, so $\sqrt{ks^2/\beta} = \sqrt{2}$. With the magnetic field further increasing, the two TSSs would decouple, which critical field is marked as $B_D$. At this time, $s_{valley} = 2$, $s_{spin} = 1$ and $s_{Kramers} = 1$, so $s = 2$. Also, $\beta = 2$ and $k = 1$, we would have the consequence $\sqrt{ks^2/\beta} = \sqrt{2}$. Secondly, the statistics law of the situation for the coupling TSSs and the $E_F$ far away from the DP can be inferred similarly, as shown in Figure 7(b).

**Table 2.** UCF amplitude with two TSSs coexisting (Unit: $c_d e^2/h$)

| TSSs | Case | $B = 0$ | $B_c < B < B_D$ | $B > B_D$ |
|---|---|---|---|---|
| Coupling | Near DP | 4 | $\sqrt{2}$ | $\sqrt{2}$ |
|  | Far away DP | 2 | $\sqrt{2}$ | $\sqrt{2}$ |
| Decoupling | Near DP | 2 | $\sqrt{2}$ | $\sqrt{2}$ |
|  | Far away DP | 1 | $\sqrt{2}$ | $\sqrt{2}$ |

Thirdly, the decoupling TSSs situation with $E_F$ near DP should be considered (Figure 7(c)). At $B = 0$, $s_{valley} = 2$ and $s_{spin} = 1$ due to the independent TSSs. The system then is GSE ($\beta = 4$) for the symmetry broken. Due to the Kramers degeneracy, $s = 4$. Also $k = 1$, so $\sqrt{ks^2/\beta} = 2$. When $B > B_c$, the TRS broken causes the system to be GUE ($\beta = 2$). The Kramers degeneracy was lifted. Also $s_{valley} = 2$ and $s_{spin} = 1$, so $s = 2$. With $k = 1$, $\sqrt{ks^2/\beta} =$ $\sqrt{2}$. Obviously, stronger magnetic field wouldn't cause additional effect. Similarly, the situation for the $E_F$ far away from DP can be easily inferred (Figure 7(d)).

These four situations are listed in Table 2. There are several features for the UCF amplitude. (1) No matter the TSS is coupled or not and wherever the $E_F$ is, once the TRS is broken, the $\delta G_{rms}$ takes the same value ($\sqrt{2}$). (2) Wherever the $E_F$ is, when the TRS is kept, the $\delta G_{rms}$ will reduce by 2 times with the TSSs changing from coupling to decoupling. (3) No matter the TSS is coupled or not, as long as the TRS exists, the UCF amplitude will reduce by 2 times with the $E_F$ moving away from DP. (4) When the TSSs are in the following two situations, UCF cannot distinguish them: (i) the TSSs are coupled and $E_F$ is far away from DP, (ii) the TSSs are decoupled and $E_F$ is near DP. In these two situations, $\delta G_{rms}$ is the same, and it will reduce $\sqrt{2}$ times with magnetic field increasing. (5) Once the two TSSs are coupled, the $\delta G_{rms}$ will reduce $2\sqrt{2}$ times with magnetic field increasing, when the $E_F$ is near DP. (6). If the TSSs are decoupled, $\delta G_{rms}$ is the same, and it will increase $\sqrt{2}$ times with magnetic field increasing, when the $E_F$ is away from DP.

### 5.3 Experimental progress

Figure 8 shows the transport behavior of a 42 nm-thick $Bi_2Te_2Se$ nanoribbon. The *R-T* curve shows the non-metallic characteristic (Figure 8(a)). Figure 8(b) shows the backgate dependent conductance, i.e. *G-V*$_G$ curve, at *B* =0 and *T* = 30 mK. There is no maximum for the resistance within this bakegate voltage range. That's to say the DP doesn't occur in the backgate voltage from -60 to 30 V. But the CF can be seen clearly. By subscribing the smooth background, the CF patterns ($\delta G$-$V_G$) are given in Figure 8(d). The fluctuation is too severe to distinguish the peak and valley features of the CF. To confirm this irregular CF is the UCF effect, we measured the CF at different temperatures, as shown in Figure 8(c). The large CF peaks are repeated at these temperatures. We enlarge a small range of the CF patterns, as shown in Figure 8(e). The $\delta G$-$V_G$ curves in the range from -4 to 2 V exhibit clear CF peaks and valleys. And they are repeatable with temperature changing. We also measured the $\delta G$-$V_G$ curves with different backgate voltage scanning directions. And the similar CF patterns are also observed, which is not shown here. All of these illustrate the repeatability of the CFs in Figure 8(b-e). It can be seen as the 'fingerprint' for the impurities in the sample. The $\delta G_{rms}$ is calculated and shown in Figure 8(f). $\delta G_{rms}$ decreases with temperature increasing, which rules out the possibility of thermal noise. Therefore, the backgate voltage induced CF is UCF effect cause by quantum interference.

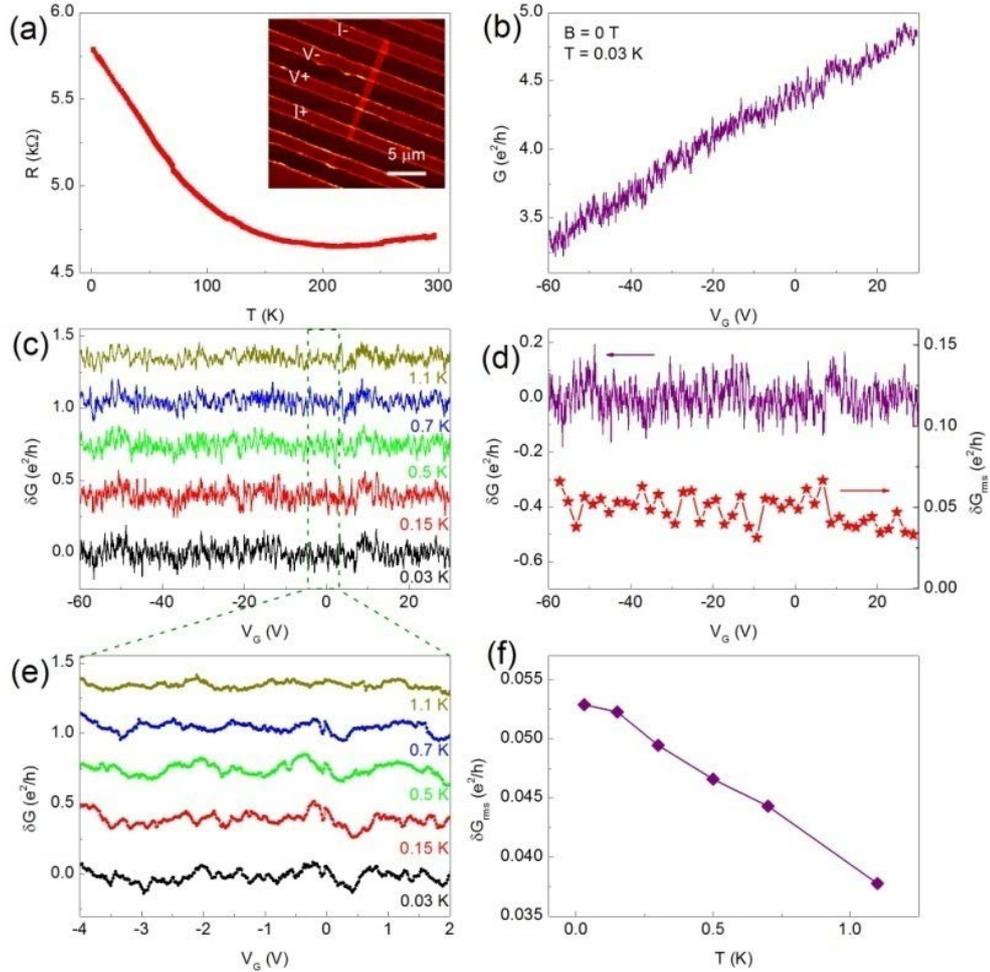

Figure 8. Tuning UCF by a gate voltage. (a) The R-T curve, the inset is the AFM image. (b) The $G$-$V_G$ curve at $T = 0.03$K and B=0T. (c) The $\delta G$-$V_G$ curves at various temperatures. The enlarged curves in the $V_G$ range -4~2V are shown in (e). (d) The $\delta G$-$V_G$ and $\delta G_{rms}$-$V_G$ curves at $T = 0.03$K. (f) Temperature dependence of $\delta G_{rms}$.

After the CF origin is confirmed, we reconsider the UCF induced by $V_G$ in Figure 8(d). If we the backgate voltage range as $\Delta V = 2$ V, and calculate the $\delta G_{rms}$ within $\Delta V$, the relationship between $\delta G_{rms}$ and $V_G$ can be seen in Figure 8(d) (the red star line). From Table 1, $\delta G_{rms}$ will reduce by 2 times with $E_F$, i.e. $V_G$, moving away from DP. In Figure 8(d), $\delta G_{rms}$ keeps unchanged in the $V_G$ range from -60 to 10 V ($\delta G_{rms} \approx 0.05\ e^2/h$). And it tends to decrease when $V_G>10$ V. Especially, at $V_G=30$ V, $\delta G_{rms}$ is already close to 0.025 $e^2/h$, i.e. it reduces 2 times. It means that the range of $V_G = -60\sim10$ V is near DP, while the range of $V_G > 10$ V is the region that far away from DP. However, it is contrary to experiment facts clearly. R-T curve gives the active gap is about 1 meV, which means $E_F$ is close to the bottom of the conduction band. On the other hand, it can be inferred that DP is in the range of $V_G < -60$ V from Figure 8(b). So actually, $V_G = -60\sim10$ V is a region that far away from DP. Then the decrease of $\delta G_{rms}$ cannot be explained by the lift of Kramers degeneracy. The true reason might be the bulk contribution cannot be neglected when $V_G>10$ V.

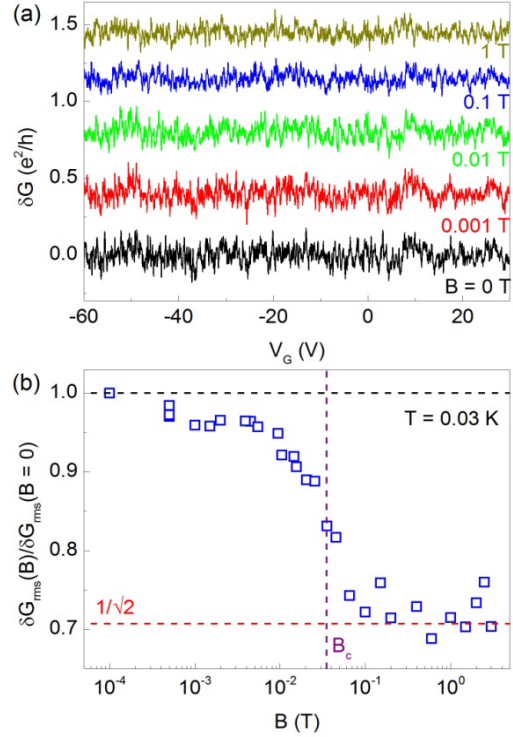

Figure 9. Tuning UCF by field at T=0.03K. (a) The $\delta G$-$V_G$ curves at various magnetic fields. (b) The magnetic field dependence of $\delta G_{rms}$.

Figure 9(a) shows the $\delta G$-$V_G$ curves at different perpendicular magnetic field. Calculating the $\delta G_{rms}$ of every curve can get the relation between $\delta G_{rms}$ and magnetic field, shown in Figure 9(b). The correlation field is exhibited to be $B_c = 0.035$ T. It can be seen that $\delta G_{rms}$ reduces $\sqrt{2}$ times with magnetic field, which can be considered to be the result of TRS broken. From Table 1, this result should be corresponding to the situation of $E_F$ near DP. But we have clarified that the experiment is in the region of $E_F$ away from DP. This is contradictory. Moreover, the result in Figure 9(b) satisfies the trivial 2DEG with strong SOC. But we have already confirmed that the 2DEG in this system cannot cause such a remarkable UCF effect in the previous section. So it cannot be explained by the statistics law of a single TSS. Two TSSs coexistence is needed.

Actually, according to the analysis of WAL, the coherence length of bulk carriers is larger than the thickness of the sample. That's to say the two TSSs are coupled with each other by the medium of bulk carriers[15]. Also, from Table 2, in the situation of $E_F$ away from DP, the theoretical prediction is that the magnetic increase will cause $\delta G_{rms}$ reduced by $\sqrt{2}$ times. This is consistent with the experiments.

Therefore, our experiments show the statistics law of the UCF in the situation TSSs coupled. The result confirms the conclusion that the topological features would disappear as well if TSSs are coupled. The decoupled situation is still in progress.

### 6. Other progress of UCF in TIs

After the introduction of CF in large bulk TI crystal, Section 4 and 5 review our group's progresses in the study of UCF in TI. Besides, there are some other reports on UCF in Tis by other groups. This section gives a brief overview of these works.

In theoretical study[46, 47, 49, 72, 73], Adroguer et al.[46] discussed the diffusion transport of TSSs. In terms of the feature of UCF, their results are like normal metals. Due to the lack of spin degeneracy for TSSs, its UCF amplitude is smaller than normal metals. Rossi et al.[49] studied the quantum transport of Dirac materials in long rang disorder, such as graphene and TIs. They inferred the relation between UCF amplitude and system shape (different ratio of $L/W$) as equation (6). But on some certain conditions, for example, near of far away from DP, the numerical results deviate from equation (6) clearly. In their results, there is no singular property for UCF of TSSs. But Zhang et al.[47] found that average conductance $\langle G \rangle$ quantized as specific fraction values in heavy disorder, $\langle G_{zz} \rangle = 1, \langle G_{xx} \rangle = 4/3, \langle G_{zx} \rangle = 6/5$, where subscript *zz* represents the current direction, and their unit is $e^2/h$. $\delta G_{rms}$ for the above three configurations are 0.54, 0.47, 0.50 $e^2/h$ respectively. However, it's difficult to measure these quantized UCF. In experiment, the measured conductance includes all the configurations, including *zz*, *xx*, *zx*. The comprehensive result may not be the quantized conductance. On the other hand, bulk and trivial 2DEG will always contribute some conductance. It's hard to separate the pure TSSs' contribution.

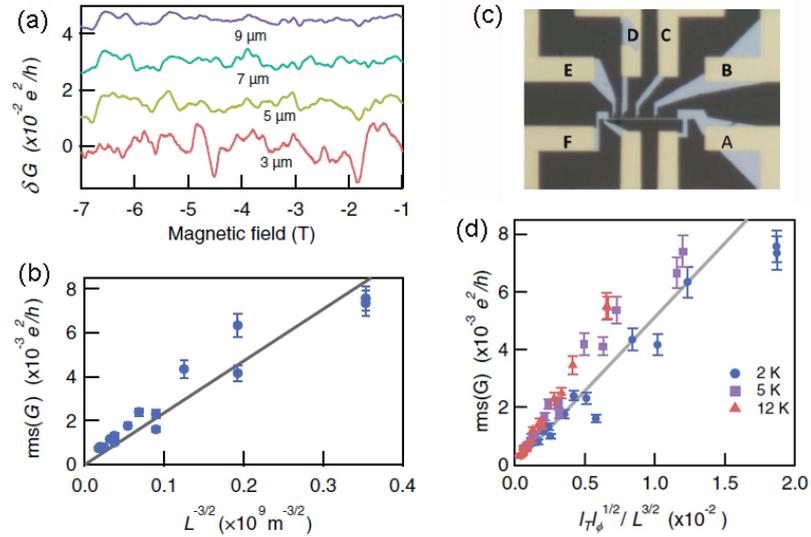

Figure 10. UCF features of a $Bi_2Se_3$ quasi-one-dimensional nanostructure. (a) The $\delta G$-$B$ curves of different sample length. (b) The relationship between $\delta G_{rms}$ and sample length L. (c) Optical image of the $Bi_2Se_3$ wire sample, the width of the whole image is 50μm. (d) $\delta G_{rms}$ as a function of $L_T L_\varphi^{1/2}/L^{3/2}$. Adapted from Ref. 77.

In the experiment study[29, 30, 74-77], Xia *et al.*[74] observed time dependent UCF in $Bi_{1.5}Sb_{0.5}Te_{1.8}Se_{1.2}$ microflake. The reason is the scattering centers (defect, impurity and so on) of the sample drift with time, or impurity states jump between the ground state and the metastable state, which causes the change of phases of electron interference change. Matsuo *et al.*[77] studied the scaling law of the UCF in $Bi_2Se_3$ quasi-one-dimension structure, as shown in Figure 10(c), it is the $Bi_2Se_3$ thin film grew by MBE. The experiment result is shown in Figure 10(a). According to the one dimension (1D) UCF theory[16], when $L_T \ll L_\phi \ll L$, the UCF amplitude can be expressed by

$$\delta G_{\text{rms}} = C \frac{e^2}{h} \left( \frac{L_T L_\phi^{1/2}}{L^{3/2}} \right), \tag{8}$$

where constant $C$ is $\sqrt{\pi/3} = 1.023\ldots$ in the condition with SOI and without TRS. The relationship between $\delta G_{rms}$ and $L^{-3/2}$ is shown in Figure 10(b). It is a linear relation, which demonstrates that the CF in $Bi_2Se_3$ quasi-1D structure satisfies the law of 1D UCF. To reveal the nature of the UCF, Figure 10(d) shows the scaling law of $\delta G_{rms}$ and $L_T L_\phi^{1/2}/L^{3/2}$. By linear fitting, the constant $C$ in equation (8) can be determined in experiment. It gives $C = 0.51\pm0.02$, which is nearly 2 times smaller than the theoretical value of 1.02. They suggest that the prerequisite, $L_T \ll L_\phi \ll L$, may not always be satisfied. On the other hand, the special transport of TSSs may cause the deviation from the traditional UCF theory. Overall, this experiment exhibits the unusual behaviors of UCF in TIs, which provides new ideas for studying.

## 7. Conclusion and outlook

We mainly review our group's progresses in the study of UCF in TIs. The results can be summed up as following. We observe the irregular CF in MC of $Bi_2Te_2Se$ nanoribbon at low temperature in the high magnetic field region. By changing the scanning direction of the field and temperature, the CF is demonstrated to be inherent. The temperature dependence rules out the possibility of thermal noise. Then it can be considered to be UCF effect. The positions of CF peaks vary with the angle by $1/\cos\theta$ law, when the angle between magnetic field and the normal of the nanoribbon is changed. This confirms the 2D signature of the UCF. This is naturally associated with 2D SSs. However, although the bulk is not metallic, there is still a small amount of residual bulk carrier with low mobility. From the WAL effect with magnetic field parallel to the axis of the nanoribbon, the coherence length of the bulk carriers is in the same order with sample thickness. If the bulk state contributes to the UCF, the quasi-2D bulk carriers would show the quasi-2D UCF features. This will bother the judgment of the origin of UCF. In consideration of the sharp dependence of UCF amplitude on the coherence length, we applied an independent parallel magnetic field to the axis of the nanoribbons to suppress the coherence of bulk carriers. By scanning the perpendicular magnetic field, the acquired UCF signal is independent to the parallel magnetic field. Then it is confirmed that the UCF is not from the bulk state, but the SSs. But other experiments report that there is trivial 2DEG in the TI surface, which could also exhibit the UCF effect. A direct idea is that the coherence of trivial 2DEG without the protection of TRS cannot do better than TSSs, so the final UCF should be contributed by the TSSs. To prove this idea, we find that the UCF amplitude of TSS is different from trivial 2DEG. Hence, we study the UCF in a series of samples, and get the inherent UCF amplitude by self-averaging scaling law. Comparing with the theoretical prediction, our experimental results are more likely that the UCF originates from the TSSs instead of trivial 2DEG. Moreover, we study the statistics law of the UCF in $Bi_2Te_2Se$ nanoribbon. From the gate voltage dependent CF at different magnetic field, we find the $\delta G_{rms}$ reduces $\sqrt{2}$ times with magnetic field incresing, which is consistent with the theory.

From these works, we find the UCF induced by TSS. Many works confirm the UCF originates from the TSS, but more obvious and convincing evidences are still in expectation. Especially, the effect of the π Berry phase of TSS on UCF is not lucid. And the influence of interaction between multiple conduction channels on UCF need to be studied further by experiment. Although there are so many works need to be done, the UCF has its unique and intriguing features on the study of TI transport.

By reviewing the progress on UCF in TI, we can find the work on this aspect is still weak. More theory and experiment works is needed. In the future study on UCF, we think the most important work is to prove the topological feature of UCF in TI. As introduced in the introduction, transport experiments, such as SdH oscillation, WAL and AB oscillation, can reveal the π Berry phase of TSS. That's why there is so much attraction on the phenomena of TI. Therefore, only by proving the uniqueness of UCF in revealing the π Berry phase in this system, the strong vitality of the UCF in the study of TI can be guaranteed. To achieve this goal, it's necessary to carry out in-depth theoretical exploration and the sophisticated designed experiment synchronously.

Note that this article is translated from Reference 78.

**Acknowledgements**

Project supported by the State Key Development Program for Basic Research of China (Grant No. 2013CB922103, 2011CB922103，2014CB921103), the National Natural Science Foundation (NSF) of China (Grant Nos. 91421109, 11023002, 11134005, 61176088), and the NSF of Jiangsu province (Grant No. BK20130054) .